\documentclass{emulateapj}
\usepackage{apjfonts}
\usepackage{amsmath,graphicx}

\shortauthors{T. Hosokawa \& K. Omukai}
\shorttitle{Protostellar Evolution of Low-metallicity Stars}

\begin{document}

\title{Low-Metallicity Protostars and the Maximum Stellar Mass \\ 
Resulting from Radiative Feedback}

\author{Takashi Hosokawa\altaffilmark{1,2} and 
        Kazuyuki Omukai\altaffilmark{2}}

\altaffiltext{1}{Jet Propulsion Laboratory, California Institute
of Technology, Pasadena CA 91109 ; takashi.hosokawa@jpl.nasa.gov}
\altaffiltext{2}{Division of Theoretical Astronomy, 
National Astronomical Observatory, Mitaka, Tokyo 181-8588, Japan;
hosokawa@th.nao.ac.jp, omukai@th.nao.ac.jp}

\begin{abstract}
The final mass of a newborn star is set at the epoch when the mass accretion
onto the star is terminated. We study the evolution of accreting protostars
and the limits of accretion in low metallicity environments.
Accretion rates onto protostars are estimated via the temperature evolution 
of prestellar cores with different metallicities.
The derived rates increase with decreasing metallicity, from
$\dot{M} \simeq 10^{-6}~M_\odot/{\rm yr}$ at $Z = Z_\odot$ to
$10^{-3}~M_\odot/{\rm yr}$ at $Z = 0$.
With the derived accretion rates,
the protostellar evolution is numerically calculated.
We find that, at lower metallicity, the protostar has
a larger radius and reaches the zero-age
main-sequence (ZAMS) at higher stellar mass.
Using this protostellar evolution,
we evaluate the upper stellar mass limit where the mass accretion
is hindered by radiative feedback.
We consider the effects of radiation pressure exerted
on the accreting envelope, and expansion of the H~II region.
The mass accretion is finally terminated by radiation pressure
on dust grains in the envelope for $Z \gtrsim 10^{-3}~Z_\odot$
and by the expanding H~II region for lower metallicty.
The mass limit from these effects increases with decreasing
metallicity from $M_* \simeq 10~M_\odot$ at $Z = Z_\odot$
to $\simeq 300~M_\odot$ at $Z = 10^{-6}~Z_\odot$.
The termination of accretion occurs after the central star
arrives at the ZAMS at all metallicities, which allows us
to neglect protostellar evolution effects in discussing
the upper mass limit by stellar feedback.
The fragmentation induced by line cooling in
low-metallicity clouds yields prestellar cores with masses large
enough that the final stellar mass is set by the feedback effects.
\end{abstract}


\section{Introduction}

What determines the mass of a newborn star?
Our knowledge on this fundamental question in astrophysics
is still limited.
Observationally, the stellar initial mass function (IMF)
is a power law with the Salpeter slope $-2.3-2.6$
on the high-mass side and becomes flatter below
a break at $\simeq 1 M_{\sun}$ on the lower-mass side.
Towards even lower mass ($\la 0.1M_{\sun}$),
the number of stars appears to decline (e.g., Muench et al. 2002).
That is, the characteristic mass in the solar neighborhood
is $\simeq 0.1 - 1~M_{\sun}$.

Recently, the mass distribution of dense cores in a star-forming
region was found to have a similar shape to the stellar IMF,
with a mass offset about a factor of three (Alves et al. 2007).
Also, the typical mass of the cores $M_{\rm core}$
is approximately the Jeans mass $M_{\rm J}$:
\begin{equation}
M_{\rm core} \sim M_{\rm J} \simeq 2~M_\odot
 \left( \frac{n}{10^5~{\rm cm}^{-3}} \right)^{-1/2}
 \left( \frac{T}{10~{\rm K}} \right)^{3/2},
\label{eq:mc_solar}
\end{equation}
where we used the typical number density
$n \simeq 10^5~{\rm cm}^{-3}$
and temperature of the cores $T \simeq 10$~K.
These facts indicate that the characteristic stellar mass is set
already at the formation stage of dense cores through
gravitational fragmentation.
The factor of three reduction of the stellar mass from the core mass
can be a result of some stellar feedback, e.g., protostellar outflows,
which reduce the amount of material in the core,
or binary/multiple formation.
Theoretical estimations support such an efficiency
for the low-mass cores (e.g., Matzner \& McKee 2000).

Even though the characteristic stellar mass is
determined by the gravitational (or turbulent) fragmentation
process, formation of the most massive stars
will be limited by the feedback from the protostar,
as inferred from the following consideration.
A small protostar, which forms at the center of the core subsequent
to its gravitational collapse, grows in mass by accretion
from the leftover core material, i.e., envelope.
Since a massive protostar becomes so luminous during accretion,
radiation pressure on dust grains easily disrupts the
accretion envelope
(Larson \& Starrfield 1971; Kahn 1974; Wolfire \& Cassinelli 1987).
Under a typical accretion rate, i.e., a Jeans mass $M_{\rm J}$ of gas
falling on the protostar in a free-fall time $t_{\rm ff}$,
\begin{equation}
\dot{M} \sim   \frac{M_{\rm J}}{t_{\rm ff}} \sim \frac{c_s^3}{G}
       \simeq 2 \times 10^{-6}~M_\odot/{\rm yr}
              \left( \frac{T}{10~{\rm K}} \right)^{3/2} ,
\label{eq:mdot_sol}
\end{equation}
where $c_s$ is the sound speed in the natal
core and $G$ is the gravitational constant,
the upper mass limit by radiative pressure
falls on about $10-20M_{\sun}$.

Although rare, stars more massive than this limit do exist in the Milky Way.
How such stars formed is still debated.
Accretion through a circumstellar disk can somewhat alleviate the
radiative feedback.
Besides, it appears that massive star formation requires
a non-standard environment.
For example, faster accretion than the standard
or stellar mergers are sometimes invoked.
In fact, observations tell us that accretion rates on massive protostars
are quite high, $\dot{M} \gtrsim 10^{-4}~M_\odot/{\rm yr}$ (e.g., Zhang
et al. 2005; Keto \& Wood 2006; Grave \& Kumar 2009),
and the natal dense cores are unusually massive
$M_{\rm core} \gtrsim 10~M_\odot \sim 10~M_{\rm J}$
(e.g., Saito et al. 2006; Motte et al. 2007).
However, as a guide, the upper mass limit within the standard
accretion scenario is still important.

Metallicity governs the thermal balance in prestellar cores,
thereby affecting the core temperature and the protostellar accretion rate
through equation (\ref{eq:mdot_sol}).
The accretion rate in turn alters the protostellar evolution.
Metals also act as catalysts in the CN cycle, modifying the stellar
structure.
In addition, lowering the dust opacity is expected to reduce
the strength of the radiative feedback.
As a first step, we here aim to calculate the evolution of
accreting protostars and evaluate the mass limit
where the accretion is prohibited by the stellar radiative
feedback as a function of metallicity
under the standard accretion rates and spherical symmetry.
We consider the effects of radiation pressure exerted on
the accretion envelope, as well as expansion of an H~II region.

The organization of this paper is as follows.
In \S~\ref{sec:pev}, we calculate the evolution of protostars
at various metallicities.
Next in \S~\ref{sec:limits}, we examine the radiative feedback
effects. We also consider the limit of mass accreted over the main
sequence lifetime.
The final stellar mass is eventually derived
as a function of the metallicity.
In \S~\ref{sec:corem}, we also compare the obtained final
stellar masses with the typical core masses at each metallicity.
\S~\ref{sec:concl} is devoted to conclusions.

\section{Protostellar Evolution with Various Metallicities}
\label{sec:pev}

\subsection{Accretion Rates onto Protostars}

\begin{figure}
\begin{center}
\includegraphics[width=3.5in]{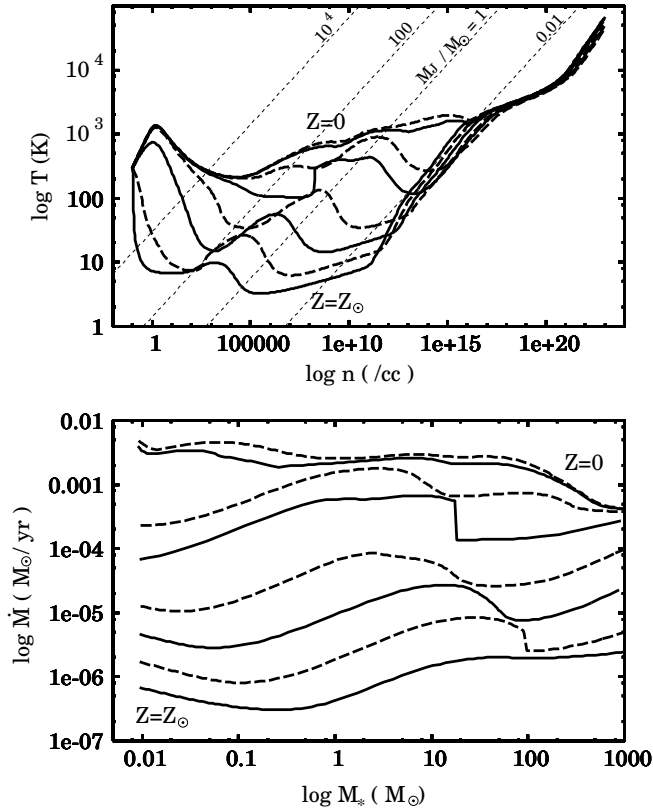}
\end{center}
\caption{
{\it Top panel:} Temperature evolution of prestellar cores with different
metallicities (taken from Omukai et al. 2005).
The Jeans mass is constant along the thin dotted lines,
$M_{\rm J} = 10^{-2}$, $1$ and $10^2$, and $10^4~M_\odot$.
In calculating $M_{\rm J}$, the hydrogen is assumed to be fully
atomic (molecular) for $M_{\rm J} \ge 10^{2}M_{\sun}$
($M_{\rm J} \le 1M_{\sun}$, respectively).
{\it Bottom panel:} The fiducial accretion rates derived
from the thermal evolution for different metallicities (see text).
In both panels, the solid and dashed curves alternately
represent the cases with
metallicities $Z =0$, $10^{-6}$, $10^{-5}$,
$10^{-4}$, $10^{-3}$, $10^{-2}$, $10^{-1}$ and $Z_\odot$.
}
\label{fig:thev_mdotz}
\end{figure}
\begin{figure*}
\begin{center}
\epsscale{0.9}
\plotone{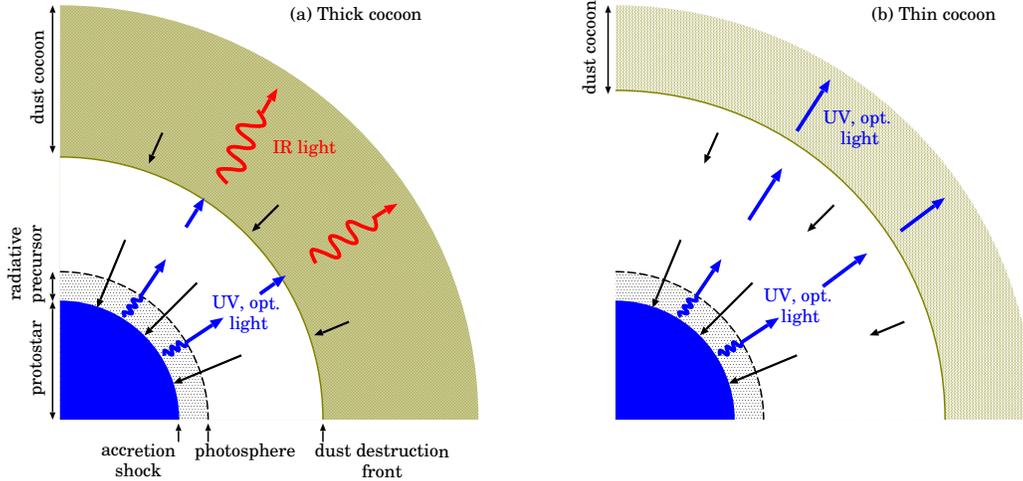}
\caption{
Schematic view of a protostar and its surrounding accretion envelope.
The accretion flow hits the stellar surface at the
accretion shock front.
If the flow becomes opaque before hitting the stellar surface,
a photosphere emerges outside the stellar surface.
The optically thick part of the flow is called a radiative precursor.
Dust grains in the envelope evaporate at a dust destruction front.
The layer outside the destruction front is a dust cocoon.
With the optically thick cocoon (left figure), most of the UV and optical
light is absorbed by grains just outside the dust destruction
front and is re-emitted as infrared light.
With the thin cocoon (right figure), on the other hand,
the stellar UV and  optical light directly escapes out
of the cocoon.}
\label{fig:env_struct_schem}
\end{center}
\end{figure*}
\begin{figure}
\begin{center}
\includegraphics[width=3.5in]{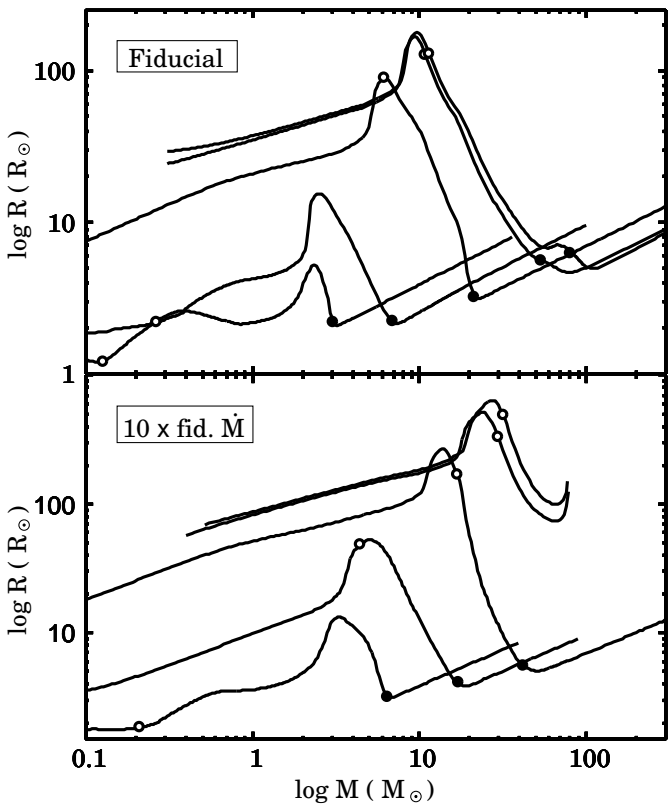}
\end{center}
\caption{
Evolution of protostellar radii with metallicities
$Z = 0$, $10^{-6}$, $10^{-4}$, $10^{-2}$ and $Z_\odot$.
The upper and lower panels show the cases with the
fiducial and ten times higher accretion rates, respectively.
The open circles on the curves mark the epoch when the energy
production rate by deuterium burning reaches 80~\% of the steady
burning rate, $L_{\rm D,st} \equiv \dot{M} \delta_{\rm D}$, where
$\delta_{\rm D}$ is the energy available from
the deuterium burning per unit gas mass.
The filled circles on the curves indicate the epoch of ZAMS arrival
defined as when the total energy production rate by hydrogen burning
reaches 80~\% of the stellar luminosity $L_*$.}
\label{fig:m_r_fid_10times}
\end{figure}

The accretion rate onto the protostar is determined
by the prestellar temperature (eq. \ref{eq:mdot_sol}),
which depends on metallicity.
The temperature evolution during the prestellar collapse for various
metallicities has been studied by Omukai et al. (2005)
by using a one-zone model with detailed
chemical and radiative processes.
The prestellar core is assumed to undergo run-away collapse,
i.e., its central flat part has a radius about the Jeans length
and the central density increases at a free-fall timescale.
Note that its radius and mass generally decreases with increasing
density and thus most of the material originally in the core is
left behind after some moment.
We adopt the result by Omukai et al. (2005) in calculating
the accretion rate, and their Figure 1
is reproduced in the upper panel of Figure \ref{fig:thev_mdotz}.
In general, prestellar temperatures are higher for lower metallicity.

The thermal evolution track $(\rho, T)$ is converted to
the radial density or temperature profile of the envelope
($r = \lambda_{\rm J}, \rho$ or $T$), where
$\lambda_{\rm J}$ is the Jeans length.
When materials within a radius $r$ have accreted on the protostar,
the protostellar mass is $M_* \simeq M_{\rm J}$,
where $M_{\rm J}$ is the Jeans mass with $(\rho, T)$ at radius $r$.
The accretion rate to this protostar is $\sim c_s^3/G$,
where $c_s$ is the sound speed at radius $r$.
We derive accretion rates as a function of the instantaneous
protostellar mass by converting the thermal evolution track
$(\rho, T)$ to $(M_* = M_{\rm J}, \dot{M} = c_s^3 / G)$.
The lower panel of Figure \ref{fig:thev_mdotz} shows those
accretion rates.
The higher accretion rates at lower metallicities
reflect the higher temperatures in the envelopes.

Here, we have adopted the standard accretion rate
$\dot{M} \sim c_s^3 / G$.
However, the accretion rate can vary according to
the way of prestellar collapse.
For example, in a dynamical runaway collapse,
the Larson-Penston (LP) solution, the accretion rate reaches
as high as $\dot{M} \simeq 47~c_s^3 / G$,
while the standard rate is almost exact for the static collapse
of the singular isothermal sphere, the Shu solution.
In realistic hydrodynamical calculations, the prestellar collapse
asymptotically approaches the LP solution.
The accretion rate is initially as high as $\sim 10$ times
the standard rate and decreases gradually until it reaches
the standard rate (e.g., Larson 2003).
To accommodate possible such variations, we also examine
the evolution with the accretion rate
ten times higher than the standard value.

\subsection{Result of Protostellar Evolution}

Next, we calculate the protostellar evolution with the derived
accretion rates at each metallicity.
Stellar feedback effects, which possibly alter the accretion rates,
are separately examined in \S~\ref{ssec:fdbk}
with the calculated evolution.
Details of our numerical method are summarized in our previous
papers (e.g., Hosokawa \& Omukai 2009 and references therein).
We explain a brief outline of the method and some improvements.
Figure \ref{fig:env_struct_schem} schematically depicts the
structure of a protostar and accretion envelope.
Dust grains in the accretion envelope evaporate at a dust
destruction front, where their temperature reaches about 2000~K.
The envelope is divided into inner gas envelope and outer dust cocoon,
whose boundary is the dust destruction front.
We numerically solve the structure of an accreting protostar and
gas envelope.
The basic equations for the protostar are the stellar
structure equations with mass accretion
(e.g., Stahler, Shu \& Taam 1980, Palla \& Stahler 1991).
We adopt the free-fall flow for the optically thin gas envelope.
When the flow becomes opaque before reaching the stellar surface,
we solve the structure of the opaque part, called a radiative precursor.
The flow finally hits the stellar surface and forms an accretion
shock front.
The whole structure of the protostar and gas envelope is
consistently determined to satisfy the shock jump conditions.
An evolutionary calculation begins with a small initial
protostar constructed according to Stahler, Shu \& Taam (1980).
We construct sequence of models with increasing stellar mass
by accretion, until the protostar reaches the zero-age
main-sequence (ZAMS) stage.
Nuclear burning reactions of deuterium, hydrogen,
and helium are included in our calculations.
Further evolution in the post main-sequence stage is out of
scope of this paper. In \S~\ref{ssec:lifet}, we roughly
estimate when the MS stage is over with mass accretion.
In this paper, we use the tabulated equation of state generated
following Pols et al. (1995).
Opacity tables at $0 < Z < Z_\odot$ are also generated by
logarithmic linear interpolation between the tables at $Z = 0$
and $Z = Z_\odot$.

Figure \ref{fig:m_r_fid_10times} presents the calculated protostellar
evolution at each metallicity.
The evolution at $Z = 0$ agrees with
previous calculations by Omukai \& Palla (2001, 2003).
There are four evolutionary stages with the fiducial
accretion rates (upper panel):
(i) adiabatic accretion ($M_* \lesssim 6~M_\odot$),
(ii) swelling ($6~M_\odot \lesssim M_* \lesssim 10~M_\odot$),
(iii) Kelvin-Helmholtz (KH) contraction
($10~M_\odot \lesssim M_* \lesssim 100~M_\odot$),
and (iv) accretion to ZAMS star ($M_* \gtrsim 100~M_\odot$).
A key quantity to understand varieties of the evolution is
a ratio between two timescales:
the accretion timescale,
\begin{equation}
t_{\rm acc} = \frac{M_*}{\dot{M}},
\end{equation}
and the KH timescale,
\begin{equation}
t_{\rm KH} = \frac{G M_*^2}{R_* L_*}.
\end{equation}
The two timescales $t_{\rm acc}$ and $t_{\rm KH}$ can be regarded
as evolutionary and thermal-adjustment timescales of the protostar.
The protostar gains materials with high entropy generated
at the accretion shock front on $t_{\rm acc}$.
Meanwhile, radiation diffuses outward and reduce the stellar
entropy on $t_{\rm KH}$.
In the early phase, $t_{\rm acc}$ is much shorter than $t_{\rm KH}$.
The entropy taken into the star adiabatically accumulates
in the stellar interior (adiabatic accretion).
This stage is over because $t_{\rm KH}$ decreases
($L_*$ increases) as the stellar mass increases.
As $t_{\rm KH}$ approaches $t_{\rm acc}$, the interior entropy is
gradually transferred outward by radiation.
A part of transferred entropy is received by the outer layer
near the stellar surface. This causes the swelling of the star.
Finally, $t_{\rm acc}$ becomes longer than $t_{\rm KH}$.
The star loses energy by radiation, and contracts
hydrostatically (KH contraction).
The central temperature increases with the KH contraction.
When hydrogen burning is ignited, the protostar
reaches the ZAMS stage.
The above four evolutionary stages can be applied for
other $Z > 0$ cases.
The protostellar radius gradually increases in the early stage
and then experiences the swelling. The protostar contracts in the next
stage and finally arrives at the ZAMS.
In all cases, the timescale balance switches from
$t_{\rm acc} < t_{\rm KH}$ to $t_{\rm acc} > t_{\rm KH}$
at the swelling.
However, Figure \ref{fig:m_r_fid_10times} also presents some
different aspects of the evolution with different metallicities.
The following three evolutionary tendencies
result from the lower accretion rates at higher metallicities
(see also Hosokawa \& Omukai 2009).

\begin{figure*}
\begin{center}
\epsscale{0.9}
\plotone{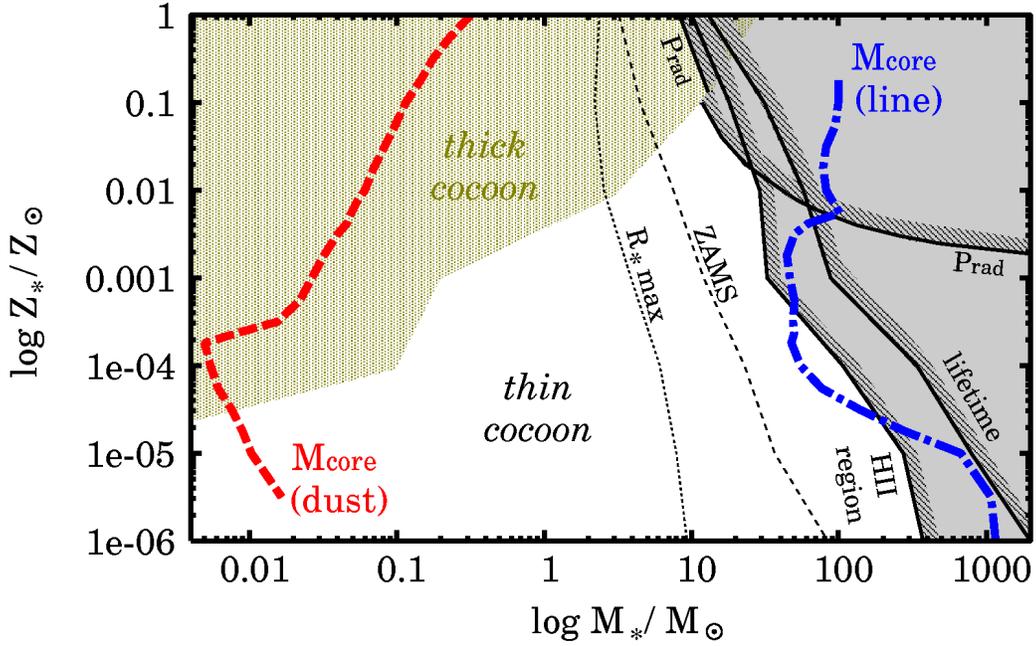}
\caption{
The stellar upper mass limit by radiative feedback
at the fiducial accretion rate as a function of metallicity.
The thin dotted and dashed lines shows two characteristic epochs
in protostellar evolution; the maximum of the stellar
radius (labeled as '$R_{\ast}$ max') and arrival to the ZAMS('ZAMS').
The dust cocoon is optically thick in the upper-left hatched
area and thin in the rest.
The right gray-shaded areas denote the region where the mass
accretion is prohibited either by the radiation pressure on the dust
cocoon ('$P_{\rm rad}$'), the expansion of an H~II region ('HII region'),
or the stellar lifetime ('lifetime').
The characteristic mass-scales of prestellar cores by fragmentation
are also presented.
The thick-dashed and dot-solid lines represent the cores produced as
a result of fragmentation induced by dust cooling and line cooling,
respectively.} 
\label{fig:tau_fid}
\end{center}
\end{figure*}

(i) The protostellar radius is larger for lower metallicity.
For $Z = 0$, the radius exceeds $100~R_\odot$ at $Z = 0$, while
only a few $R_\odot$ for $Z = Z_\odot$.
This reflects higher entropy content in the star
for higher accretion rate (see e.g., Hosokawa \& Omukai 2009)
as the radius is related with entropy $s$ by
\begin{equation}
R_{\ast} \propto M_{\ast}^{-1/3}{\rm exp}[const. \times s].
\end{equation}
Without enough time after the shock for accreted matter
to cool radiatively before burying into opaque interior,
the interior entropy is higher for higher accretion rate.
(ii) The protostellar mass is higher at the onset of ZAMS phase for
lower metallicity. The star arrives on the ZAMS
at $\simeq 100~M_\odot$ for $Z = 0$, and at
a few $M_\odot$ for $Z = Z_\odot$ (also see Fig.\ref{fig:tau_fid}).
Since the stellar radius is larger for lower metallicity
(e.g., lower accretion rate), the interior temperature
is higher according to the relation,
\begin{equation}
T = \frac{\mu}{\cal{R}} \frac{P}{\rho}
 \propto \frac{M_{\ast}}{R_{\ast}}.
\label{eq:t_typ}
\end{equation}
Thus, higher $M_{\ast}$ is needed for ignition of hydrogen burning
at $\sim 10^7$~K.
(iii) Signatures of deuterium burning appear only at
metallicity $Z > 10^{-2} Z_\odot$.
At $Z = Z_\odot$, the protostar becomes almost fully convective
by deuterium burning at $M_* \simeq 0.1~M_\odot$. The linear
increase of the radius at
$0.1~M_\odot \lesssim M_* \lesssim 0.3~M_\odot$ is
due to this effect. After that, the star gradually returns to
be radiative by the swelling at $M_* \simeq 2~M_\odot$.
This is in contrast to the metal-poor cases.
At $Z = 0$, deuterium burning begins later at $M_* \simeq 10~M_\odot$
and hardly affects the evolution. The protostar remains radiative
until its arrival to the ZAMS.
These variations are also due to different evolution of the
stellar inner temperature following equation (\ref{eq:t_typ}).

Apart from different accretion rates, metallicity still
influences the protostellar evolution.
For example, the accreting MS star has smaller radius
at the lower metallicity.
This is due to different initial abundances of C and N atoms.
Massive MS stars is supported against gravity
with hydrogen burning via CN-cycle reactions.
With lower CN abundance, higher temperature is required to
supply enough to support the star, thus leading to smaller MS radius.

The lower panel of Figure \ref{fig:m_r_fid_10times} shows the
protostellar evolution with accretion rate ten times higher
than the fiducial value.
The mass-radius relation varies with metallicity in the same
manner as in the fiducial cases.
Compared to the fiducial case at the same metallicity,
a protostar has larger radius and arrives to the ZAMS
at higher stellar mass, which is consistent with
the dependence on accretion rates discussed above.
Clear differences are found only for $Z = 0$ and $10^{-6}~Z_\odot$.
The protostar turns to inflate during the KH contraction in these
cases. We separately focus on these cases in
\S~\ref{sssec:prad_prc} below.

\section{Limits of Mass Accretion}
\label{sec:limits}

\begin{figure*}
\begin{center}
\epsscale{0.9}
\plotone{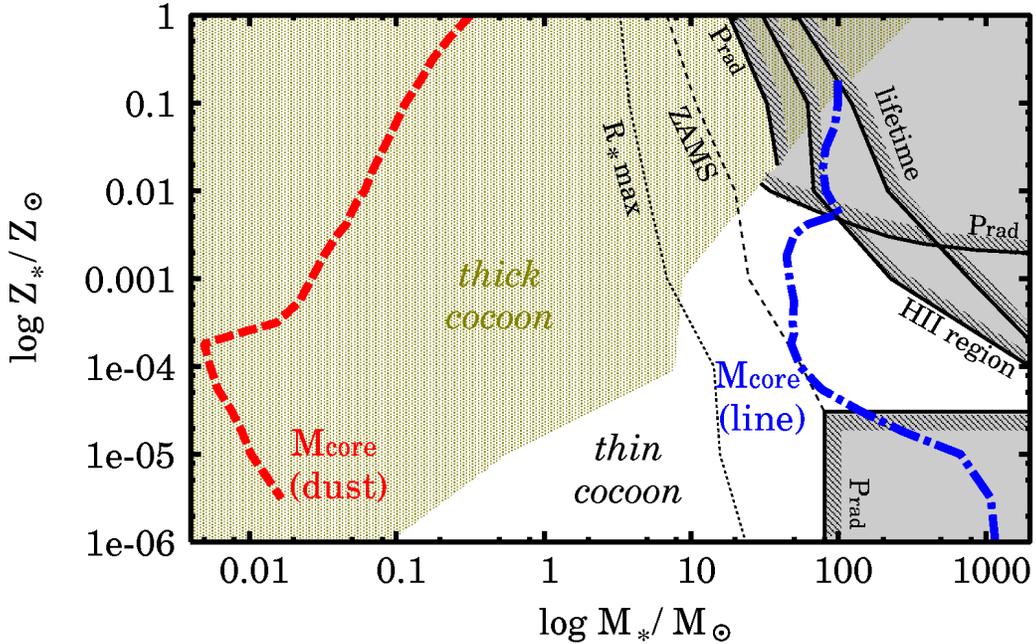}
\caption{Same as Fig.\ref{fig:tau_fid} but for the accretion rate
ten times higher than the fiducial value.
The right-bottom gray-shaded rectangular area denotes the region
where mass accretion is prohibited by the radiation pressure exerted
on the radiative precursor.
} 
\label{fig:tau_10times}
\end{center}
\end{figure*}
\begin{figure}
\begin{center}
\includegraphics[width=3.5in]{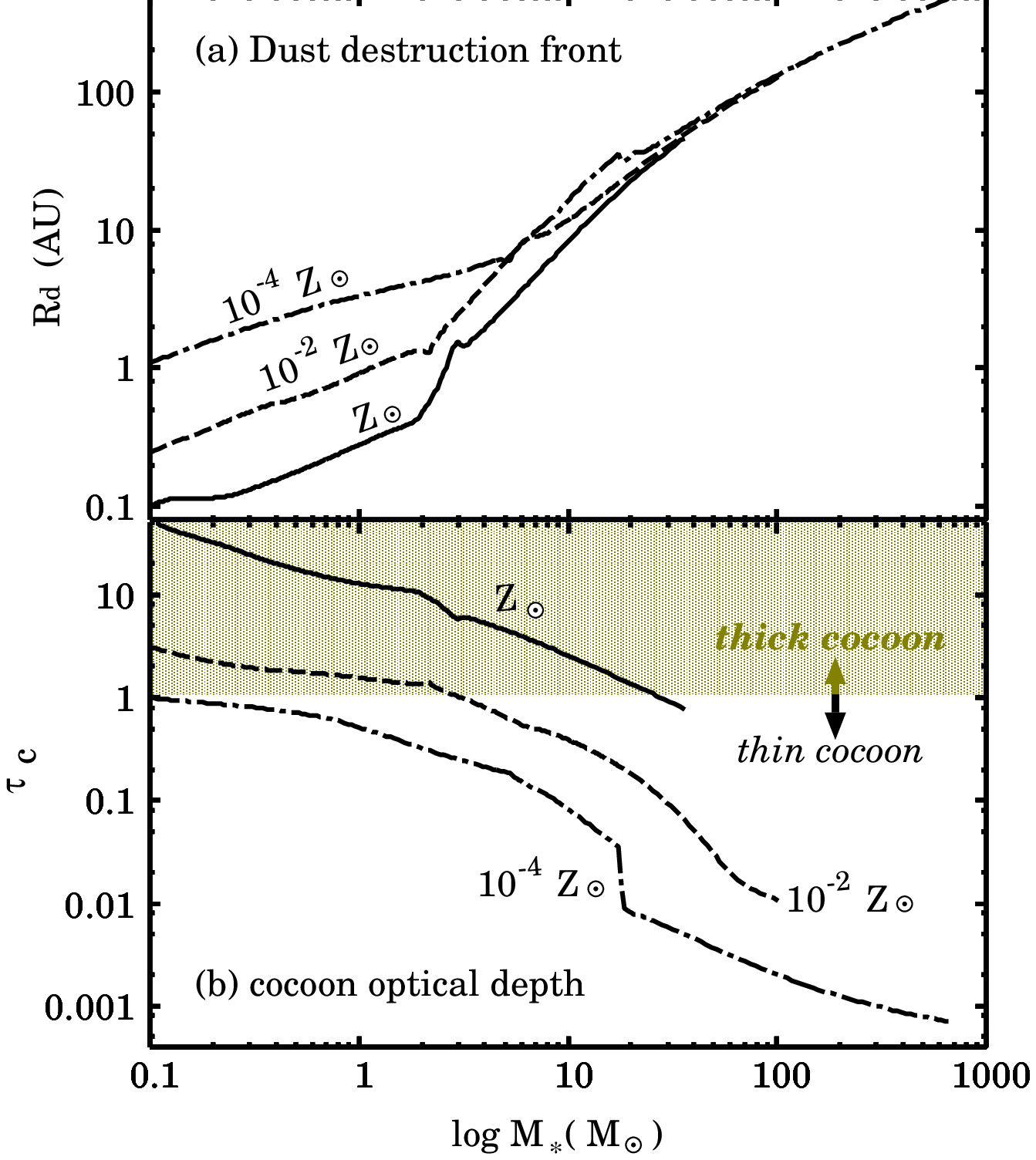}
\end{center}
\caption{
Evolution of the dust cocoon at different
metallicities; $Z = Z_\odot$ (solid lines), $10^{-2}~Z_\odot$
(dashed lines), and $10^{-4}~Z_\odot$ (dot-dashed lines).
The fiducial accretion rates are adopted at each metallicity.
{\it Upper panel:} Position of the dust destruction front, i.e.,
the inner edge of the dust cocoon.
{\it Lower panel:} Optical depth of the cocoon
for the stellar UV and optical light $\tau_{\rm c}$.
The cocoon is optically thick if $\tau_{\rm c} > 1$,
which is indicated by the dotted background.}
\label{fig:env_struct}
\end{figure}

\subsection{Radiative Feedback Effects}
\label{ssec:fdbk}

Protostellar luminosity increases with growth of the
star by accretion.
Radiative feedback effects will eventually terminate the accretion.
Below we examine the effects of radiation pressure exerted
on a radiative precursor and dust cocoon, as well as
of the expansion of an H~II region.
Our calculations have already included radiation pressure on
the radiative precursor.
We examine the other effects by separate analytic treatments and
evaluate when these effects limit the accretion
as a function of metallicity.

\subsubsection{Radiation Pressure on Radiative Precursor}
\label{sssec:prad_prc}

A protostar growing with very high accretion rate exerts
strong radiation pressure on a radiative precursor, which is
a optically thick part of the gas envelope.
This effect is found in the calculated protostellar evolution
with $10\dot{M}_{\rm fid}$ for $Z \leq 10^{-5}~Z_\odot$
(the lower panel of Fig. \ref{fig:m_r_fid_10times}).
In these cases, at $M_* \simeq 60~M_\odot$
the KH contraction turns to the abrupt expansion
(also see Omukai \& Palla 2003).
The expansion occurs because radiation pressure decelerates
the accreting flow before reaching the stellar surface.
The stellar surface pressure, which balances the ram pressure by
the flow,  becomes too low to maintain the
hydrostatic structure of the star.
The expansion occurs when the total stellar luminosity $L_{\rm tot}$
approaches the Eddington limit
$L_{\rm Edd} \equiv 4 \pi c G M_* / \kappa_{\rm e}$, where
$\kappa_{\rm e}$ is the Thomson scattering opacity.
Although the luminosity from stellar interior $L_{\ast}$ remains below
the Eddington limit, the total luminosity $L_{\rm tot}$ from the protostar
has an extra contribution from the accretion shock:
\begin{equation}
L_{\rm tot} = L_* + L_{\rm acc}
\equiv L_* + \frac{G M_* \dot{M}}{R_*} .
\end{equation}
Note that the internal to accretion luminosity ratio
$L_*/L_{\rm acc}$ is equal to the timescale ratio $t_{\rm acc}/t_{\rm KH}$.
Since $L_{\rm acc} \propto \dot{M}/{R_*}$,
$L_{\rm acc}$ becomes very high and $L_{\rm tot}$ reaches the Eddington limit
for a high enough value of $\dot{M}$
during the KH contraction, where the radius approaches the ZAMS value.
This happens for the accretion rate exceeding
the critical value
$\dot{M}_{\rm cr} \simeq 4 \times 10^{-3}~M_\odot/{\rm yr}$
for the case of $Z=0$ and depending on metallicity only weakly
(Omukai \& Palla 2003; Hosokawa \& Omukai 2009).
This is indeed the case for $Z \leq 10^{-5}~Z_\odot$
with $10\dot{M}_{\rm fid}$.
In such cases, the protostars cannot keep accreting at
its original rate after the abrupt expansion.
If the accretion terminates at this moment, the final
stellar mass will be $M_* \simeq 80~M_\odot$
(Fig. \ref{fig:tau_10times}).

\subsubsection{Radiation Pressure on Dust Cocoon}

We here estimate the effect of radiation pressure on the dust cocoon.
First, we consider how the radiation propagates outward
in the accretion envelope.
Most of the stellar radiation is initially emitted in
UV and optical wavelengths.
The stellar radiation travels freely between the photosphere
and dust destruction front.
Subsequent outward propagation in the dust cocoon
depends on whether the cocoon is optically thick or thin
for the stellar UV/optical light.
If the cocoon is optically thick,
most of the direct light is absorbed near the dust destruction front
and is re-emitted in infrared wavelengths,
then propagates outward diffusively.
This is the case for the present-day protostars still embedded in
their natal dense cores, which are observed as bright infrared sources.
With lower abundance of dust grains, however, the cocoon
becomes transparent even for the stellar UV/optical light.
In this case, the direct stellar light escapes from the cocoon
without absorption and re-emission.
Figure \ref{fig:env_struct_schem} schematically shows the
structure of the flow for these two cases.
The optical depth of the cocoon is roughly estimated as
\begin{equation}
\tau_{\rm c} \sim \kappa_{\rm UV/opt}
                   \rho_{\rm d}
                   R_{\rm d},
\label{eq:tauc}
\end{equation}
where $\kappa_{\rm uv \cdot opt}$ is dust opacity for optical
and UV light, and the suffix ``d'' denotes physical quantities at
the dust destruction front. We regard the cocoon as optically thin
(thick) if $\tau_{\rm c} < 1$ $(>1, respectively)$.
The optical depth $\tau_{\rm c}$ can be evaluated as follows.
The position of the dust destruction front is
\begin{equation}
R_{\rm d} \simeq 35~{\rm AU}
\left( \frac{L_{\rm tot}}{10^5~L_\odot} \right)^{1/2}
\label{eq:rd}
\end{equation}
(Wolfire \& Cassinelli 1987). The upper panel of Figure
\ref{fig:env_struct} shows evolution of the dust destruction front
at $Z = 10^{-4}$, $10^{-2}$, and $Z_\odot$ for the fiducial accretion
rates. The front moves outward with increasing stellar mass as
the total luminosity $L_{\rm tot}$ increases.
We adopt a typical dust opacity in the optical range
(e.g., McKee \& Ostriker 2007),
\begin{equation}
\kappa_{\rm uv \cdot opt}
\simeq 200~{\rm cm^2/g} \left( \frac{Z}{Z_\odot}  \right) .
\end{equation}
For a free-falling flow, the density at $R_{\rm d}$ is
\begin{equation}
\rho_{\rm d} = \frac{\dot{M}}{4 \pi \sqrt{2 G M_* R_{\rm d}^3} } .
\label{eq:rhod}
\end{equation}
We derive
$\tau_{\rm c} \propto \rho_{\rm d} R_{\rm d} \propto L_{\rm tot}^{-1/4}$
by substituting equations (\ref{eq:rd}) and (\ref{eq:rhod})
into (\ref{eq:tauc}).
Hence the optical depth decreases with the increase of stellar
mass and luminosity.
Even if the cocoon is initially optically thick, it becomes
optically thin at some moment in the evolution.
Evolution of the optical depth shown in lower panel
of Figure \ref{fig:env_struct} demonstrates this transition.
At $Z = 10^{-2}~Z_\odot$, for example, the cocoon becomes transparent
for the direct light at $M_* \simeq 3~M_\odot$.
Note that the optical depth $\tau_{\rm c} \propto \kappa_{\rm UV/opt}
\rho_{\rm d}\propto Z \dot{M}$ does not linearly scale with $Z$,
owing to another dependence on $\dot{M}$.
Figures \ref{fig:tau_fid} and \ref{fig:tau_10times} summarize
the evolution of the dust cocoon for all cases.

Next, we consider the condition for the flow to overcome the radiation
pressure for cases where the cocoon is optically thick or thin,
respectively.
In the optically thick case,
most of the direct radiation is absorbed in a thin layer near
the dust destruction front and isotropically re-emitted subsequently
in infrared wavelengths.
Hence the absorbing layer receives outward momentum flux of
$\frac{L_{\rm tot}}{4 \pi R_{\rm d}^2 c}$.
For the accretion to proceed, the inward momentum flux of the flow
must exceed this outward flux (Wolfire \& Cassinelli 1987),
\begin{equation}
\rho_{\rm d} u_{\rm d}^2 > \frac{L_{\rm tot}}{4 \pi R_{\rm d}^2 c} ,
\label{eq:pram}
\end{equation}
where $u_{\rm d}$ is the free-fall infall velocity,
\begin{equation}
u_{\rm d} = \sqrt{\frac{2 G M_*}{R_{\rm d}}} .
\label{eq:ud}
\end{equation}
Using the continuity equation $\dot{M} = 4 \pi R^2 \rho u$, equation
(\ref{eq:pram}) leads to a condition for accretion rates,
\begin{equation}
\dot{M} > \frac{L_{\rm tot}}{c u_{\rm d}} .
\label{eq:pram2}
\end{equation}
Additionally, the flow in the cocoon must overcome
the radiation pressure by the infrared re-emission.
This condition can be stated as that
the stellar total luminosity be less than the
Eddington limit defined with dust opacity
for infrared light $\kappa_{\rm IR}$:
\begin{equation}
L_{\rm tot} < L_{\rm Ed, IR}
           \equiv \frac{4 \pi c G M_*}{\kappa_{\rm IR}} .
\label{eq:ledir}
\end{equation}
Otherwise, radiation pressure decelerates the accretion
flow before reaching the dust destruction front.
The infrared opacity in the local star-forming regions takes
its maximum at $T_{\rm dust} \simeq 600$~K (e.g., Pollack et al. 1994).
We adopt the Rosseland mean opacity at this maximum as the infrared
opacity
\begin{equation}
\kappa_{\rm IR} \simeq 8~{\rm cm^2/g} \left( \frac{Z}{Z_\odot}  \right) .
\end{equation}
On the other hand, for an optically thin cocoon,
the stellar radiation is not converted to infrared radiation by absorption
and re-emission.
The only condition is that the total stellar
luminosity be less than the Eddington limit for
the opacity at the UV/optical light $\kappa_{\rm uv \cdot opt}$,
\begin{equation}
L_{\rm tot} < L_{\rm Ed, uv \cdot opt}
           \equiv \frac{4 \pi c G M_*}{\kappa_{\rm uv \cdot opt}} .
\label{eq:leduv}
\end{equation}

Figures \ref{fig:tau_fid} and \ref{fig:tau_10times} show
the limit when the radiation pressure becomes so strong to violate
the above conditions.
For $Z = Z_\odot$ with the fiducial accretion rate, for example,
the flow is optically thick until $M_* \simeq 30~M_\odot$.
For an optically thick flow, equation (\ref{eq:pram2}) usually
put the stricter condition than equation (\ref{eq:ledir}), and
the upper limit falls on $M_* \simeq 8~M_\odot$ for
$Z = Z_\odot$ (Figure \ref{fig:tau_fid}).
At $Z = 10^{-2}~Z_\odot$, the optically thick/thin transition
of the flow occurs at $\simeq 3M_{\sun}$ and the accretion
continues until $M_* \gtrsim 50~M_\odot$, where
the condition (\ref{eq:leduv}) is violated.
Both figures show that the limiting mass increases with decreasing
the metallicity. Radiation pressure on the dust cocoon does not
limit mass accretion at $Z \lesssim 10^{-3}~Z_\odot$.

\subsubsection{Expansion of an H~II Region}

\begin{figure}
\begin{center}
\includegraphics[width=3.5in]{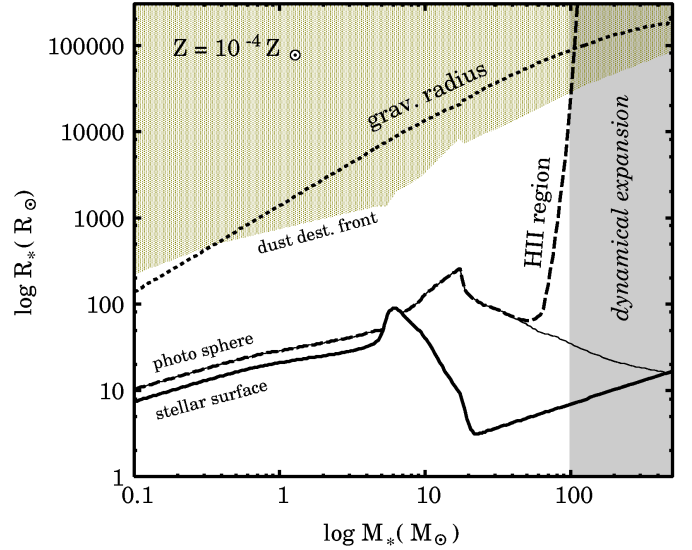}
\end{center}
\caption{
Evolution of the protostar and surrounding H~II region
at the fiducial accretion rate for metallicity $10^{-4}~Z_\odot$.
The thick and thin solid lines represent the position of the stellar surface,
i.e., the accretion shock front, and the photosphere within the accreting flow.
Positions of the ionization front $R_{\rm HII}$ and
gravitational radius for an ionized gas $R_{\rm g}$ are presented
with dashed and dotted lines.
The steady mass accretion is disturbed by dynamical
expansion of the H~II region when $R_{\rm HII} > R_{\rm g}$,
which is denoted by the gray shaded background.
The dot-dashed line represents the dust destruction front,
the region above which corresponds the dusty accreting envelope.}
\label{fig:hii_z1e-4}
\end{figure}

The stellar UV luminosity increases with its mass.
Strong UV photons ionize gas near the star and an H~II region
develops inside the accreting envelope.
For the free-falling flow with accretion rate $\dot{M}$,
the radius of the H~II region $R_{\rm HII}$ can be expressed
as a function of the UV photon number luminosity $S$:
\begin{equation}
R_{\rm HII} = R_{\rm ph} \exp \left( \frac{S}{S_{\rm cr}} \right) ,
\label{eq:rhii}
\end{equation}
where $R_{\rm ph}$ is radius of the photosphere and
the critical UV luminosity $S_{\rm cr}$ is defined as,
\begin{equation}
S_{\rm cr} = \frac{\alpha \dot{M}^2}{8 \pi \mu_{\rm H}^2 G M_*} ,
\end{equation}
where $\alpha$ is the recombination efficiency and $\mu_{\rm H}$ is
mass per hydrogen atom.
As long as $S<S_{\rm cr}$, the HII region is confined near the stellar
surface.
Once the stellar UV luminosity exceeds $S_{\rm cr}$, the HII
region grows exponentially and soon encloses all materials
in the natal core.
Figure \ref{fig:hii_z1e-4} presents the evolution of the protostar
and its H~II region for $Z = 10^{-4}~Z_\odot$ with the fiducial
accretion rate.
In this case, the photosphere locates beyond the accretion shock
front at the stellar surface except short duration around
$M_{\ast}\simeq 6M_{\sim}$.
The UV photon number luminosity is calculated by
\begin{equation}
S = 4 \pi R_{\rm ph}^2
   \int_{\nu_{\rm L}}^{\infty} \frac{B(\nu, T_{\rm eff})}{h \nu}
   d\nu ,
\end{equation}
where $\nu_{\rm L}$ is the Lyman-limit frequency, $h$ is
Planck constant, and $B(\nu, T_{\rm eff})$ is Planck function
with the effective temperature $T_{\rm eff}$.
Figure \ref{fig:hii_z1e-4} shows that the H~II region begins to expand
exponentially at $M_* \simeq 40~M_\odot$.

The evolution of the HII region and the accretion flow experiences
the following two stages, depending on the ratio of the
radius of the HII region $R_{\rm HII}$ and the gravitational
radius of an ionized gas,
\begin{equation}
R_{\rm g} \equiv \frac{G M_* \phi_{\rm Ed}}{c_{\rm HII}^2},
\end{equation}
where $c_{\rm HII}$ is the sound speed of ionized gas and
$\phi_{\rm Ed} \equiv 1 - L_{\rm tot}/L_{\rm Ed}$ is a correction
factor taking account of repulsive force by radiation pressure.
When $R_{\rm HII} < R_{\rm g}$, the accreting flow supersonically
passes through the ionization front.
Although the temperature and pressure jump up behind the
ionization front, this hardly influences the flow dynamics.
The heated gas inside the H~II region is also gravitationally
bound to the central protostar.
Once $R_{\rm HII}$ exceeds $R_{\rm g}$,
the accreting flow becomes subsonic and dynamical
expansion of the H~II region disrupts the accreting envelope.
The gas inside the H~II region begins to move outward
owing to the pressure excess of the ionized gas.
A shock front emerges ahead the ionization front, and
sweep up materials into a shell around the H~II region.

Strictly speaking, the density distribution of the flow
deviate from the free-fall one even without rotation.
Radiation pressure via photoionization decelerates the flow
and increases the density. Omukai \& Inutsuka (2002) showed that
this effect can significantly delay the dynamical expansion of
an H~II region around the first star ($Z=0$).
However, McKee \& Tan (2008) argued that
relaxing the spherical symmetry reduces this effect.
In the rotating core collapse, density in the polar region
decreases by a large factor by the epoch when an H~II region forms.
As a result, radius of the H~II region reaches $R_{\rm g}$
before the trapping effect becomes important
(see \S~5.2 of Mckee \& Tan 2008 for detail).
The trapping effect becomes even weaker for lower accretion rates,
i.e., for higher metallicity.
In this paper, although we treat the spherically symmetric flow
in estimating the radiation feedback, small asphericity
is assumed to exist to alleviate the trapping effect by
the photoionization.

Figure \ref{fig:env_struct} shows that the dynamical expansion
occurs at $M_* \simeq 100~M_\odot$ at $Z = 10^{-4}~Z_\odot$.
The stellar mass limit by the HII region expansion is summarized
in Figures \ref{fig:tau_fid} and \ref{fig:tau_10times} as a
function of metallicity.
With the fiducial accretion rates, the expansion of an H~II region
is the primary feedback effect disrupting the accreting envelope for
$Z \lesssim 10^{-3}~Z_\odot$.
The mass limit increases from $M_* \simeq 30~M_\odot$ to $300~M_\odot$
with decreasing metallicity.
This is because higher accretion rate can quench the H~II region of
the more massive star.
With ten-times the fiducial accretion rate, the mass limit exceeds
$1000~M_\odot$ for $Z \lesssim 10^{-4}~Z_\odot$.
In these cases, not the expansion of an HII region, but
the radiation pressure on the radiative precursor
limits mass accretion for $Z \lesssim 10^{-5}~Z_\odot$
(\S~\ref{sssec:prad_prc}).

The mass limit by the H~II region expansion is
$M_* \simeq 300~M_\odot$ for $Z=0$ with the fiducial accretion rates.
This happens to be comparable to that
by McKee \& Tan (2008), $M_* \simeq 100 - 300~M_\odot$.
However, they consider mass accretion via an accretion disk formed by
rotating core collapse. The breakout of an H~II region beyond $R_{\rm g}$
occurs earlier at $M_* \simeq 50 - 100~M_\odot$. Photoevaporation of the
accretion disk finally quenches mass accretion.
The same process can work in halting the accretion also for
finite metallicities, and should be explored in future work.

\subsection{Mass Limit by Stellar Lifetime}
\label{ssec:lifet}

We have shown that mass accretion to a protostar is limited
by radiative feedback effects, i.e., radiation pressure exerted on a
dust cocoon for $Z \gtrsim 10^{-3}~Z_\odot$, and expansion of
an H~II region for lower metallicity.
Figures \ref{fig:tau_fid} and \ref{fig:tau_10times}
show that these limits are reached after the protostar arrives at the ZAMS.
For very high feedback limits,
accreting stars might end their lives
before reaching such limits.
The maximum mass of the star can gain in its lifetime is
roughly given by
\begin{equation}
M_{\ast} \simeq \int^{t_{\rm MS}(M_{\ast})}_{0} \dot{M}dt
\end{equation}
where $t_{\rm MS}$ is the lifetime of a ZAMS star
of $M_*$.
This limit is also presented in Figures
\ref{fig:tau_fid} and \ref{fig:tau_10times}.
Here, we adopt $t_{\rm MS}$ of primordial stars from Marigo et al.
(2001) for $M_* \leq 100~M_\odot$ and from Schaerer (2002) for
$M_* \geq 100~M_\odot$, as their dependence on metallicity is weak.
We see that the limit by the lifetime is
less stringent than those by the radiative feedback effects.
Hence the final stellar mass will be set by the feedback effects,
as long as the prestellar core is massive enough.

\section{Characteristic Core Mass Scales}
\label{sec:corem}

The mass limit derive above is reached only in the case that
the mass reservoir, i.e., the mass of the natal core,
is sufficient.
Here, we discuss the characteristic mass of the prestellar cores
and compare them with the mass limits by the feedback.
The cores are formed as a result of fragmentation of
a more massive cloud.
According to analytical (Schneider et al. 2002; Larson 2005)
and numerical (Bromm, Coppi \& Larson 2002; Jappsen et al. 2005;
Clark, Glover \& Klessen 2008; Smith et al. 2009) arguments,
fragmentations likely occur around temperature minima of
thermal evolution tracks (see Figure \ref{fig:thev_mdotz}),
and cores of about the Jeans mass $M_{\rm J}$ are produced.
This figure shows that the tracks at
$10^{-5}~Z_\odot \lesssim Z \lesssim 0.1~Z_\odot$
experiences two local temperature minima, i.e.,
one at lower density $10^{2-5}{\rm cm^{-3}}$,
where $M_{\rm J} \simeq 100-1000~M_\odot$,
and the other at higher density $10^{5-14}{\rm cm^{-3}}$,
where $M_{\rm J} \simeq 0.01-0.1~M_\odot$.
The former is induced by the line cooling,
i.e, cooling by the H$_2$ lines for $\la 10^{-3}Z_{\sun}$, and
the metal fine-structure lines for higher metallicity,
while the latter is by the dust cooling.
In clouds with extremely low-metallicity,
dust-induced minimum disappears, thus only high-mass
fragmentation is allowed.
On the other hand, at $Z \simeq Z_\odot$, two minima
almost merges and result in a single characteristic mass-scale.
The typical core mass in the cases
with two fragmentation epochs is not clear yet.
We speculate that massive and small cores corresponding to two
fragmentation epochs exist simultaneously, and their number ratio
depends on the efficiency of dust-induced fragmentation.

Those characteristic core mass-scales are also shown in
Figure \ref{fig:tau_fid} and \ref{fig:tau_10times} for
comparison with the stellar mass limits by the feedback.
The characteristic core masses by the dust-induced fragmentation
is much smaller than the stellar mass limits.
In such cores, the final stellar masses are bounded by
the amount of mass reservoir rather than the stellar feedback.
On the other hand, the typical core masses by the line-induced fragmentation
are comparable to the mass limit by the radiative effects.
In such massive prestellar cores, therefore,
the final stellar masses should be limited by the radiative effects
and very massive stars can be formed in low-metallicity environments.

\section{Summary}
\label{sec:concl}

We studied protostellar evolution as a function of the metallicity, 
using the results to evaluate the significance of radiative 
feedback effects.  For metallicities between zero and Solar, 
we showed how the feedback terminates the mass accretion and 
determines the final stellar mass. 
Our results are summarized as follows:

First, we estimated accretion rates onto a protostar from
the thermal evolution of a prestellar core at each metallicity.
The accretion rates are higher at the lower metallicity, ranging
from $\dot{M} \simeq 10^{-3}~M_\odot/{\rm yr}$ at $Z=0$ to
$\simeq 10^{-6}~M_\odot/{\rm yr}$ at $Z = Z_\odot$.
This is because the core temperature is higher at the lower
metallicity.

Next, we numerically calculated the protostellar evolution
with the derived accretion rates at each metallicity.
The calculated evolution depends on the metallicity.
For lower metallicity, a protostar has larger radius
and arrives at the ZAMS at higher stellar mass.
These variations reflect the difference in the accretion rates.

We also examined radiative feedback effects that may halt the
mass accretion onto a protostar, i.e., radiation pressure on a radiative
precursor and dust cocoon, and expansion of an H~II region.
The radiation pressure exerted on the dust cocoon limits mass
accretion for $Z \gtrsim 10^{-3}~Z_\odot$. The upper mass limit increases
with decreasing metallicity.
At low metallicity,
the high accretion rate and low dust opacity enable the accreting
flow to overcome the radiation pressure barrier.
For $Z \lesssim 10^{-3}~Z_\odot$,
the expansion of an H~II region instead will disrupt the accretion envelope.
The mass limit resulting from the H~II region expansion also increases
with decreasing metallicity because
higher density in the flow hinders the development of the HII region.
These feedback effects become significant after the protostar reaches
the ZAMS, but before the stellar lifetime is over.
With the accretion rates ten times higher than the
fiducial values, radiation pressure on the radiative precursor
limits the steady mass accretion
at $M_* \simeq 100~M_\odot$ for $Z \lesssim 10^{-5}~Z_\odot$.

Cores of two typical mass-scales are produced by
fragmentation over wide ranges in metallicity.
One scale comes from line cooling ($100-1000 M_{\sun}$),
and the other from dust cooling ($0.01-1 M_{\sun}$).
The mass limit due to the radiative feedback effects is far higher
than the typical masses with dust cooling, and
comparable to those with line cooling.
In massive cores formed by line-induced fragmentation,
the radiative feedback effects terminate mass accretion
and set the final stellar masses.

{\acknowledgements
The authors thank Neal Turner for his help to revise the manuscript.
This study is supported in part by Research Fellowships of the Japan
Society for the Promotion of Science for Young Scientists (TH) and
by the Grants-in-Aid by the Ministry of Education, Science and
Culture of Japan (18740117, 19047004, 21684007: KO).
}

\end{document}